\documentclass[mathleft
]{an}
\usepackage{graphicx}

\usepackage{times}
\begin{document}

% The following seven commands are intended for editorial usage and should be ignored by
% the author(s).
\Pagespan{1}{}% Document's page range. 
% If second parameter is left empty, the last page is computed automatically.
\Yearpublication{2010}%
\Yearsubmission{2010}%
\Month{07}%   
\Volume{999}%  
\Issue{88}% 
% \DOI{This.is/not.aDOI}% 

%Irregularity in the magnetic field measurement of the Of?p star HD\,108
\title{The kinematic characteristics of magnetic O-type stars}

\date{Received date / Accepted date}

\author{{S. Hubrig\inst{1}\fnmsep\thanks{Corresponding author: \email{shubrig@aip.de}}}
\and
N.V. Kharchenko\inst{1,2}
\and
M. Sch\"oller\inst{3}
}

\titlerunning{Kinematic characteristics of O stars}
\authorrunning{S. Hubrig et al.}

\institute{
Astrophysikalisches Institut Potsdam, An der Sternwarte 16, 14482 Potsdam, Germany
\and
Main Astronomical Observatory, 27 Academica Zabolotnogo Str., 03680 Kiev, Ukraine
\and
European Southern Observatory, Karl-Schwarzschild-Str.\ 2, 85748 Garching, Germany
}

%\received{}
%\accepted{}
%\publonline{later}


\abstract{
Although magnetic fields have been discovered in ten massive O-type stars during the last years,
the origin of their magnetic fields remains unknown. 
Among the magnetic O-type stars, two stars, HD\,36879 and HD\,57682, were identified as candidate runaway stars in the past,
and $\theta^1$\,Ori\,C was reported to move rapidly away from its host cluster.
We search for an explanation for the occurrence of magnetic fields in O-type stars by examining the assumption
of their runaway status.
We use the currently best available astrometric, spectroscopic, and photometric data to calculate 
the kinematical status of seven magnetic O-type stars with previously unknown space velocities. 
The results of the calculations of space velocities suggest that five out of the seven magnetic O-type stars 
can be considered as candidate runaway stars. Only two stars, HD\,155806 and
HD\,164794, with the lowest space velocities, are likely members of Sco\,OB4 and NGC\,6530,
respectively. However, the non-thermal radio emitter HD\,164794 is a binary system with colliding winds, for which
the detected magnetic field has probably a different origin in comparison to other magnetic O-type stars.
}

\maketitle

\section{Introduction}
\label{sect:intro}

Only in the last years, magnetic fields have been detected in a number of O and early B-type stars. 
To date, ten O-type stars have published magnetic fields:
$\theta^1$\,Ori\,C (Donati et al.\ \cite{Donati2002}), HD\,191612 (Donati et al.\ \cite{Donati2006}), 
HD\,155806 (Hubrig et al.\ \cite{Hubrig2007}), $\zeta$\,Ori\,A (Bouret et al.\ \cite{Bouret2008}),
HD\,36879, HD\,148937, HR\,6272, 9\,Sgr (Hubrig et al.\ \cite{Hubrig2008}),
HD\,57682 (Grunhut et al.\ \cite{Grunhut2009}),
and HD\,108 (Martins et al.\ \cite{Martins2010}, Hubrig et al.\ \cite{Hubrig2010}).
However, theories on the origin of magnetic fields in O-type stars
are still poorly developed.
This is due to the fact that the distribution of magnetic field strengths in massive stars
from the ZAMS to more evolved stages,
which would shed light on the origin of the magnetic field,
has not yet been studied.
A number of magnetic O-type stars seem to be slow rotators and exhibit an excess
of nitrogen (e.g., Walborn et al.\ \cite{Walborn2003}, Naz\'e et al.\ \cite{Naze2008a,Naze2008b}). 
Note that massive stars with such characteristics are only very rarely found in clusters 
(Trundle et al.\ \cite{Trundle2007}), contrary 
to field stars, which frequently show slower rotation speed and nitrogen enrichment
(Gies \& Lambert \cite{GiesLambert1992}). 
Wolff et al.\ (\cite{Wolff2007}) suggested that locking of accretion disks 
to proto-stars results in slowly rotating stars and these disks are longer
lived in the field population than in clusters.
Clearly, to understand the origin of magnetic fields in massive stars, it is 
important to build trustworthy statistics on their occurrence.
About 10\% of main-sequence A and B stars are slowly 
rotating, chemically peculiar, magnetic Ap and Bp stars, and among their descendants, 
the white dwarfs, 10\% have strong magnetic fields.
The magnetic fields in magnetic white dwarfs could be 
fossil remnants from the main-sequence phase, consistent with magnetic flux conservation
(Ferrario \& Wickramasinghe \cite{FerrarioWickramasinghe2005}).
If we assume that magnetic fields of massive stars behave like Ap and Bp stars, then we would expect a magnetic field 
probability of 10\%. 

The catalogue of Galactic O-stars (GOSV2; Sota et al.\ \cite{Sota2008}) contains about 370 O-type stars.
105 O-stars (28\%) in the subset of this catalogue are considered as field stars since they 
could not be identified as present or former members of recognised clusters or OB associations.
Seventy-three of these 105 stars were recently back-traced to nearby associations 
and young open clusters (Schilbach \& R\"oser \cite{SchilbachRoeser2008}).
Remarkably, almost one third of the star sample studied by 
Schilbach \& R\"oser (\cite{SchilbachRoeser2008}; corresponding to $\sim$10\% of the total number of O-stars in 
the GOSV2 catalogue) appeared to
have rather long travel  times after their ejection from their parent open clusters, 
longer by up to 60\% than the expected lifetime of O-type stars. 
We suggest that one of the possible explanations for the 
long lifetime of these runaway stars could be the presence of magnetic fields.
To test the proposition that magnetic O-type stars are
frequently runaway stars, we carried out a study of  
the kinematical status of O-type stars with detected magnetic fields.

\section{Massive stars in the field and characteristics of runaway stars}
\label{sect:runaway_characteristics}

It is generally accepted that the majority of massive stars form in star clusters or 
associations. 
To explain the origin of massive stars in the field, two mechanisms were discussed in the literature. 
In one scenario, 
close multibody interactions in a dense cluster environment cause one or more stars to be scattered 
out of the region (e.g.\ Leonard \& Duncan \cite{LeonardDuncan1990}). For this  mechanism, runaways are 
ejected in dynamical three- or four-body interactions and
the model predicts that most runaways will be single stars,
although some close binaries can be expected under exceptional circumstances.
An alternative mechanism involves a supernova explosion 
within a close binary, ejecting the secondary due to the conservation of momentum 
(Zwicky \cite{Zwicky1957}, Blaauw \cite{Blaauw1961}).
The resulting neutron star may remain bound to the secondary 
if not enough mass is lost during the explosion. 
The production of runaway massive binaries is expected to be rather low:
Portegies Zwart (\cite{PortegiesZwart2000}) predicted 
a binary fraction of 20\%--40\% among runaways that are ejected by a binary supernova scenario
using binary population synthesis calculations. 
The observed fraction of binaries among runaways seems to be consistent with either scenario 
(5\%---26\% in Mason et al.\ \cite{Mason1998}).

Schilbach \& R\"oser (\cite{SchilbachRoeser2008}) followed the dynamic history of field O-type stars,
calculating the path of stars and clusters back in time in the Galactic potential.
For this study they used the positions and proper motions from the PPMX 
catalogue (R\"oser et al.\ \cite{Roeser2008}), 
and the radial velocities from the CRVAD-2 catalogue  (Kharchenko et al.\ \cite{Kharchenko2007}).
The major source of specific information on open clusters and O-type stars were the catalogues 
of Kharchenko et al.\ (\cite{Kharchenko2005a,Kharchenko2005b}).
Acceptable solutions for the trajectories through the Galaxy were achieved for 73 O-type stars,
indicating that the present-day data are 
consistent with the assumption that the major part of O-type field stars were ejected from young open clusters 
or protoclusters over the past 10\,Myr. 
Since one third of the studied stars appear to
have rather long travel  times since their ejection from different open clusters, 
one of the possible explanations for the long lifetime of these runaway stars could be 
their rejuvenation by mass transfer just prior to the supernova explosion,  which would reset their 
effective zero-age times to an epoch just prior to ejection. 
Another explanation could be that these stars are overluminous for their mass in the same way as some 
massive X-ray binaries (Kaper \cite{Kaper2001}), or they are all rapid rotators, where fast rotation helps
to mix gas and extend their main-sequence life.
The rapid rotation, however, cannot be considered as a major characteristics of runaway stars, 
since rather long rotation periods and low $v$\,sin\,$i$-values were determined for a few runaway stars in 
individual studies (e.g.\ Grunhut et al.\ \cite{Grunhut2009}).

In our view, the paradox between the long travel distance and the apparent young age of these stars
could possibly be explained if runaway stars did undergo a rejuvenation in tight massive binary systems,
where two stars merge at the end of their interaction and somehow acquire a magnetic field during this 
process.  
As Pflamm-Altenburg \& Kroupa (\cite{PflammAltenburgKroupa2010}) showed, only a small fraction of 
runaway O-type stars (1--4\%) can be produced by the combined effect of 
massive binary ejection from star clusters and a second acceleration of a massive star during a 
subsequent supernova explosion.
We note, however, that the current parameters in observable massive binaries are insufficient to pin down
the many uncertainties about the mass-transfer process (e.g.\ Dray \& Tout \cite{DrayTout2007}), 
and how the rejuvenated star becomes magnetic.  
Maeder \& Meynet (\cite{MaederMeynet2005})
showed that the presence of a magnetic field enlarges the main-sequence lifetime by 
$\sim$10\% for a 15\,M$_\odot$ star.
However, no estimations were ever made for more massive stars.

\section{Kinematical properties of O-type stars with published magnetic fields}
\label{sect:mf_characteristics}

Among the sample of magnetic O-type stars, two stars, HD\,36879 and HD\,57682, were already identified as
candidate runaway stars 
(e.g., de Wit et al.\ \cite{deWit2004}, \cite{deWit2005}, Comeron et al.\ \cite{Comeron1998}). The space motion of the star $\theta^1$\,Ori\,C 
was studied by van Altena et al.\ (\cite{vanAltena1988}), who reported that $\theta^1$\,Ori\,C is 
moving at 4.8$\pm$0.5\,km\,s$^{-1}$
towards position angle 142$^{\circ}$ and that this velocity is significantly larger that the dispersion 
value of 1.5$\pm$0.5\,km\,s$^{-1}$ found for 
the other cluster members. The results of the radial velocity study of Stahl et al.\ (\cite{Stahl2008})
indicate that this star is moving rapidly away from the Orion Molecular Cloud and its host cluster. 

\begin{table*}
\centering
\caption{
HD numbers, stellar positions, proper motions, radial velocities, parallaxes, and magnitudes of magnetic O-type stars.
}
\label{tab:objects}
\begin{tabular}{rrrr@{$\pm$}lr@{$\pm$}lrr@{$\pm$}lr@{$\pm$}lr@{$\pm$}l}
\hline
\multicolumn{1}{c}{HD} &
%\multicolumn{1}{c}{RA 2000} &
%\multicolumn{1}{c}{Dec 2000} &
\multicolumn{1}{c}{l} &
\multicolumn{1}{c}{b} &
\multicolumn{2}{c}{PMx} &
\multicolumn{2}{c}{PMy} &
\multicolumn{1}{c}{RV} &
\multicolumn{2}{c}{$\pi$} &
\multicolumn{2}{c}{B} &
\multicolumn{2}{c}{V}\\
\multicolumn{1}{c}{number} &
%\multicolumn{1}{c}{[hour]} &
%\multicolumn{1}{c}{[deg]} &
\multicolumn{1}{c}{[deg]} &
\multicolumn{1}{c}{[deg]} &
\multicolumn{2}{c}{[mas/yr]} &
\multicolumn{2}{c}{[mas/yr]} &
\multicolumn{1}{c}{[km\,s$^{-1}$]} &
\multicolumn{2}{c}{[mas]} &
\multicolumn{2}{c}{[mag]} &
\multicolumn{2}{c}{[mag]}\\
\hline
   108 & 117.9221 &     1.2470 & $-$5.12 & 1.11 & $-$1.15 & 0.93 & $-$62.0 & $-$0.01 & 0.64 & 7.505 & 0.007 & 7.375 & 0.007 \\
 37742 & 206.4522 & $-$16.5852 &   3.99  & 0.74 &    2.52 & 0.40 & 45.5    &    3.99 & 0.79 & 1.798 & 0.005 & 1.898 & 0.004 \\
148937 & 336.3614 & $-$0.2124  &    0.60 & 1.42 & $-$5.12 & 1.15 & $-$33.1 &    0.82 & 1.30 & 7.038 & 0.010 & 6.757 & 0.009 \\
152408 & 344.0789 &     1.4969 & $-$1.21 & 1.11 & $-$2.28 & 0.82 & $-$56.5 &    0.33 & 0.73 & 5.946 & 0.004 & 5.807 & 0.008 \\
155806 & 352.5859 &     2.8683 & 0.33    & 1.00 & $-$2.02 & 0.50 &    10.9 &   0.65  & 0.76 & 5.559 & 0.004 & 5.613 & 0.005 \\
164794 &  6.0090  &  $-$1.2050 &   1.92  & 1.19 & $-$0.40 & 0.99 & $-$12.0 &   0.66  & 1.00 & 5.940 & 0.005 & 5.933 & 0.005 \\
191612 & 72.9871  &    +1.4365 & $-$3.26 & 0.96 & $-$6.65 & 1.02 &  $-$5.2 &    0.18 & 0.74 & 8.035 & 0.009 & 7.821 & 0.009 \\
\hline
\end{tabular}
\end{table*}

\begin{table*}
\centering
\caption{
Space velocities with respect to the Galactic open clusters system (SV$_{\rm C}$)
%with respect to the LSR (SV$_{\rm LSR}$),
%and with respect to the Sun (SV$_{\rm S}$),
and the corresponding Galactic velocity components.
}
\label{tab:calc}
\begin{tabular}{rc|cc|rrrr|r@{$\pm$}lr@{$\pm$}lr@{$\pm$}lr@{$\pm$}l}
\hline
\multicolumn{1}{c}{HD} &
\multicolumn{1}{c|}{Spectral} &
\multicolumn{1}{c}{M$_V$} &
\multicolumn{1}{c|}{(B$-$V)$_0$} &
\multicolumn{1}{c}{dist} &
\multicolumn{1}{c}{X} &
\multicolumn{1}{c}{Y} &
\multicolumn{1}{c|}{Z} &
\multicolumn{2}{c}{SV$_{\rm C}$} &
\multicolumn{2}{c}{U} &
\multicolumn{2}{c}{V} &
\multicolumn{2}{c}{W} \\
\multicolumn{1}{c}{number} &
\multicolumn{1}{c|}{Type} &
\multicolumn{2}{c|}{[mag]} &
\multicolumn{4}{c|}{[pc]} &
\multicolumn{8}{c}{[km\,s$^{-1}$]} \\
\hline
   108 & O6.5 V    & $-$5.3 & $-$0.32 & 2510 & $-$1175 &   2217 &    74 &  94 & 19 &    93 & 12 & $-$13 &  7 &     2 & 12 \\ % 87   18  84  11 -25   7  -4  12
 37742 & O9.7 Iab  & $-$6.5 & $-$0.24 &  391 &  $-$335 & $-$167 & $-$91 &  32 &  6 & $-$31 &  5 &  $-$7 &  2 &     2 &  1 \\ % 45    5 -41   4 -19   2  -4   1
%148937 & O7 V      & $-$5.2 & $-$0.32 & 1043 &     956 & $-$418 &    16 &  31 & 11 & $-$27 &  6 &    10 &  6 & $-$12 &  6 \\ % 41   10 -37   5  -1   6 -19   6
148937 & O6 V      & $-$5.4 & $-$0.32 & 1144 &    1048  & $-$458 &    15 &  32 & 13 & $-$26 &  5 &    8 &  7 & $-$13 &  7 \\ % 41   10 -37   5  -1   6 -19   6
152408 & O8 Iab    & $-$6.7 & $-$0.30 & 1694 &    1629 & $-$464 &    64 &  50 & 13 & $-$50 &  6 &     7 &  8 &     1 &  8 \\ % 60   12 -59   5  -4   7  -5   8
155806 & O7.5 III  & $-$5.7 & $-$0.32 & 1251 &    1239 & $-$161 &    82 &  19 &  9 &    19 &  6 &     1 &  5 &     0 &  4 \\ % 16    8  10   4  -9   4  -7   4
164794 & O5 I      & $-$7.2 & $-$0.33 & 2615 &    2600 &    273 & $-$35 &  24 & 21 &  $-$3 &  6 &    17 & 14 & $-$15 & 14 \\ % 27   20 -13   5   5  13 -22  13
191612 & O7 V      & $-$5.2 & $-$0.32 & 1874 &     548 &   1792 &    67 &  71 & 14 &    70 & 10 & $-$11 &  5 &     0 &  9 \\ % 65   13  61   8 -23   5  -7   9
\hline
\end{tabular}
\end{table*}

Using available data on stellar positions, proper motions, Hipparcos parallaxes, radial velocities, 
photometry, and cluster catalogues,
we investigated the kinematical status for the remaining seven O-type stars with detected magnetic fields. 
The values for the stellar positions, proper motions, radial velocities, parallaxes, and B and V magnitudes
in the Johnson system are presented in Table~\ref{tab:objects}. 
They have been 
retrieved from the All-Sky Compiled Catalogue of 2\,501\,313 stars 
(ASCC-2.5; Kharchenko \& R\"oser \cite{KharchenkoRoeser2009}) 
and from Kharchenko et al.\ (\cite{Kharchenko2007}).
If available, the values for radial velocities and spectral classification were 
taken from the following individual studies:
from Humphreys (\cite{Humphreys1978}) and Martin (\cite{Martins2010})
for HD\,108,
Bouret et al.\ (\cite{Bouret2008}) for HD\,37742, 
Naz\'e et al.\ (\cite{Naze2008a}) for HD\,148937,
Conti et al.\ (\cite{Conti1977a}) for HD\,152408,
Conti et al.\ (\cite{Conti1977b}) for HD\,155806,
Naz\'e et al.\ (\cite{Naze2010}) for HD\,164794,
and Howarth et al.\ (\cite{Howarth2007}) for HD\,191612.
An uncertainty of 5\,km\,s$^{-1}$ was assumed for the radial velocity determinations. 

The calculated space velocities and their Galactic rectangular components
are presented in Table~\ref{tab:calc}. Apart from the parallax value for HD\,37742, no accurate parallaxes 
have been measured for the other stars.
For this reason, we used the method of indirect estimates of distances through
the photometric approach, which was previously used by Kharchenko et al.\ (\cite{Kharchenko2005a}).
Spectral types, corresponding absolute visual magnitudes and (B$-$V)$_0$ on the ZAMS are listed in Cols.\ 2 to 4. 
The spectral class--colour--absolute magnitude calibration was based on Straizys (\cite{Straizys1992}).
The errors in M$_V$ and (B$-$V)$_0$ were assumed as 0.5 and 0.01\,mag, respectively.
The distances and rectangular galactic coordinates X, Y, and Z with respect to the Galactic plane 
are shown in Cols.\ 5--8. Space velocities with respect to the Galactic open cluster system and the corresponding
Galactic velocity components are listed in Cols.\ 9 to 12. Solar motion parameters derived with 
open cluster system, ((U,V,W)$_{\sun}$ = (9.44, 11.90, 7.20)), 
Oort's constants and $Z$, which stand for the distance of the Sun from the Galactic plane, $Z_{\sun}$ = 20 pc, 
have been determined by Piskunov et al.\ (\cite{Piskunov2006}). 

In this paper we discuss runaway magnetic O-type stars, which originated in open clusters
and OB-associations. For this reason, we consider their motions with respect to
the Galactic open cluster system. Nowadays there is an increasing number of
evidence that the open cluster system is very close in its kinematics to the local standard rest (LSR).
Recently, Sch\"onrich et al.\ (\cite{schoen2010}) re-examined the stellar kinematics of the Solar neighbourhood  
in terms of the velocity of the Sun with respect to the LSR.
They obtained Solar motion parameters ((U,V,W)$_{\sun}$ = (11.1, 12.24, 7.25)), which are very close to those
derived with open clusters.  
%The LSR represents by definition the motions of the Local Population~I. 
%Since Galactic open clusters are typical objects of Population~I and the difference between the motion of the 
%LSR and the cluster system is only random, there is no need to recalculate space velocities with respect to the LSR.}
%This makes us free from the necessity
%of recalculation of the derived velocities in the system related to the LSR.
%To examine our results for possible differences, we also used the parameters from
%Sch\"onrich et al.\ (\cite{schoen2010}) to recalculate the space velocities for the stars in our sample.
%The results of these calculations as well as the space velocities with respect to
%the Sun (SV$_{\rm S}$) are presented in the lower part of Table~\ref{tab:calc}.
%The difference between space velocities
%calculated with respect to the Galactic open cluster system and those obtained using the LSR
%from Sch\"onrich et al.\ (\cite{schoen2010}) are always less than 2\,km\,s$^{-1}$.

Blaauw (\cite{Blaauw1961}) assigned the stellar runaway status to stars with space velocities
larger than 40\,km\,s$^{-1}$. On the other hand, Stone (\cite{Stone1979}) and Tetzlaff et al.\ (\cite{tetz2010}) 
specified the velocity cutoff at 28\,km\,s$^{-1}$.
This velocity cutoff is adopted in 
the following discussion of the obtained results. The errors in the determination of space velocities
in the most distant stars are rather large because of proper motion errors.
For this reason, we refrain from 
calling the studied stars bona fide runaways, but rather candidate runaway stars.

From the membership studies of Galactic open clusters and associations, two O-type stars, HD\,155806 and
HD\,164794, with the lowest space velocities, are likely members of Sco\,OB4 and NGC\,6530,
respectively (Kharchenko et al.\ \cite{khar04}). No other star in the sample is known to belong
to an open cluster or an OB association. HD\,155806 is also classified as an Oe star, possibly representing 
the higher mass analogues of classical Be stars (e.g. Walborn \cite{Walborn1973}). Only six members
are suggested to belong to this group of stars (e.g. Negueruela et al.\ \cite{neg2004}).
The star HD\,164794 is a spectroscopic double-lined system with an orbital period of 2.4\,yr, known as
emitting non-thermal radio-emission, probably associated with colliding winds (Naz\'e et al.\ \cite{Naze2010}).
There are only about a dozen of non-thermal radio emitting O-type stars known (e.g.\ De Becker \cite{beck2007})
and the study of magnetic fields in such stars is especially difficult due to their broad and very variable
line profiles caused by wind-wind collision.

The Of?p star HD\,148937 possesses a space velocity of 32\,km\,s$^{-1}$ with respect to the Galactic
open cluster system, with the velocity component 
$U$=$-$26 directed opposite from the Galactic center and the velocity component $W$=$-$13
directed from the Galactic plane.
These rather large velocities indicate that this star can be considered as a candidate runaway star.
HD\,148937 is surrounded by the circumstellar nebula NGC\,6164-65, expanding, with a projected velocity of about 
30\,km\,s$^{-1}$ and it is assumed that this nebula has been ejected during an LBV-like event (Leitherer \& 
Chavarria \cite{Leitherer1987}).
Among the remaining four stars, HD\,108, HD\,37742 ($\zeta$\,Orionis\,A), HD\,152408, and HD\,191612, the O-type 
supergiant $\zeta$\,Orionis\,A, with the weakest magnetic field in our sample of magnetic O-type stars,
shows the lowest space velocity with respect to the Galactic open clusters: $SV_{\rm C}$=32$\pm$6\,km\,s$^{-1}$.
The longitudinal magnetic field of $\zeta$\,Orionis\,A is of the order of a few tens of G, while for all
other stars, the longitudinal magnetic field is of the order of hundreds of G.
The other three stars, the well known Of?p stars HD\,108 and HD\,191612, and the supergiant HD\,152408, are moving 
with higher space velocities, from
50\,km\,s$^{-1}$ for HD\,152408 up to 94\,km\,s$^{-1}$ for HD\,108, suggesting that all of them can be 
considered as candidate runaway stars.

\section{Discussion}
\label{sect:discussion}

The results of our kinematical analysis seem to indicate that the presence of a magnetic field 
is more frequently detected in candidate runaway stars than in stars belonging to 
clusters or associations. The peculiar 
velocities of three magnetic O-type 
stars were already mentioned in the literature, and our results of the 
calculations of space velocities suggest that five of the remaining seven magnetic O-type stars 
can be considered as candidate runaway stars. We note, however, that the sample of stars with magnetic field detections 
is still very small and a study of a larger sample is urgently needed to confirm the detected trend.   
Unfortunately, no dedicated magnetic field surveys of O stars in clusters/associations and in the field were carried out so far.
In the sample of magnetic O-type stars, the two stars HD\,155806 and HD\,164794, with the lowest space velocities,
are most probably members of Sco\,OB4 and NGC\,6530, respectively (Kharchenko et al.\ \cite{khar04}).
However, the non-thermal radio emitter HD\,164794 is a binary system with colliding winds, for which
the detected magnetic field has probably a different origin in comparison to other magnetic O-type stars.

According to several studies (e.g., Gies \cite{Gies1987}, Mason et al.\ \cite{Mason1998},
Sota et al.\ \cite{Sota2008}),
about 70\% of the massive O-type stars in the Galaxy are observed to be associated
with stellar clusters and/or OB-associations. At least one third of the remaining
30\% of the O-type stars, i.e.\ about 10\%, are runaway stars, and may therefore also have 
formed in a cluster,
where O stars can acquire high spatial velocities after dynamical interactions
or after supernova explosions in binary systems. 
This percentage of runaways stars is similar to the percentage of Ap and Bp stars among main-sequence 
A and B stars.
Possible paths for the formation of Ap and Bp stars
were recently analysed with modern theories for the evolution of single and binary stars
by Tutukov \& Fedorova (\cite{TutukovFedorova2010}), suggesting that merging of close binaries 
is the main channel for their formation. The low binary frequency of
Ap and Bp stars of about 20\% (e.g.\ Abt \& Snowden \cite{AbtSnowden1973}) and a strong deficit 
of SB2 binaries seem to support such a formation scenario. Similarly,  
the observed binary fraction of runaway O-type stars is of the order of 5\% to 26\%
(Mason et al.\ \cite{Mason1998}), i.e.\ roughly the same as that of magnetic Ap and Bp stars.
Based on this line of arguments, it is quite possible that a significant fraction of the
runaway stars possesses magnetic fields acquired during the ejection from their parent clusters or
protoclusters.

Another aspect, which may hint at the presence of a magnetic field in runaway stars 
is that a number of individual abundance studies indicate nitrogen 
enrichment in the atmospheres of runaway stars (e.g.\ Boyajian et al.\ \cite{Boyajian2005}).
Among the magnetic O-type stars in our sample, three stars, HD\,108, HD\,148937, and HD\,191612, were 
analysed by Naz\'e et al.\ (\cite{Naze2008a,Naze2008b}), who demonstrated a possible  nitrogen enhancement
in these stars too. 
The link between the anomalous abundances and the presence of magnetic fields was
recently discovered also in massive early B-type stars.
The observations collected by
Morel et al.\ (\cite{Morel2008}) highlight a higher incidence 
of magnetic fields in hot B-type stars with nitrogen  excess and boron depletion. 

{
\acknowledgements
We would like to thank the referee for his valuable comments. NVK thanks for the support by DFG grant RO 528/10-1.
We would like to thank W.~J.\ de Wit for fruitful discussions.
}

%\appendix

\label{lastpage}

\end{document}